\documentclass[aps,pra,superscriptaddress,twocolumn,showpacs]{revtex4}
\usepackage{graphics}
\usepackage{amsmath}
\usepackage{graphicx}
\usepackage{amsfonts}
\usepackage{amssymb}

\newtheorem{theorem}{Theorem}
\newtheorem{lemma}[theorem]{Lemma}
\newtheorem{corollary}[theorem]{Corollary}

\begin{document}

\title{Computable bounds for the discrimination of Gaussian states}
\author{Stefano Pirandola}
\affiliation{MIT - Research Laboratory of Electronics, Cambridge MA 02139, USA}
\author{Seth Lloyd}
\affiliation{MIT - Research Laboratory of Electronics, Cambridge MA 02139, USA}
\affiliation{MIT - Department of Mechanical Engineering, Cambridge MA 02139, USA}
\date{\today }

\begin{abstract}
By combining the Minkowski inequality and the quantum Chernoff bound, we
derive easy-to-compute upper bounds for the error probability affecting the
optimal discrimination of Gaussian states. In particular, these bounds are
useful when the Gaussian states are unitarily inequivalent, i.e., they
differ in their symplectic invariants.
\end{abstract}

\pacs{03.67.Hk, 03.65.Ta, 02.10.Ud}
\maketitle

\section{Introduction}

One of the central problems in statistical decision theory is the
discrimination between two different probability distributions, intended as
potential candidates for describing the values of a stochastic variable. In
general, this statistical discrimination is affected by a minimal error
probability $p^{(N)}$, which decreases with the number $N$ of (independent)
observations of the random variable. The general problem of determining $%
p^{(N)}$ was faced by H. Chernoff in 1952 \cite{Chernoff}. Remarkably, he
derived an upper bound, today known as \textquotedblleft Chernoff
bound\textquotedblright , having the non trivial property of providing $%
p^{(N)}$ in the limit of infinite observations (i.e., for $N\rightarrow
+\infty $). Very recently, a quantum version of this bound has been
considered in Refs.~\cite{QCbound0,QCbound1}. Such a \textquotedblleft
quantum Chernoff bound\textquotedblright\ allows estimation of the minimal
error probability $P^{(N)}$ which affects a corresponding quantum problem,
known as \emph{quantum state discrimination}. In this problem, a tester aims
to distinguish between two possible quantum states of a system, supposing
that $N$ identical copies of the system are available for a generalized
quantum measurement. The problem of quantum state discrimination is
fundamental in several areas of quantum information (e.g., quantum
cryptography \cite{QKDreview}) and, in particular, for continuous variable
quantum information \cite{CVbook}. Continuous variable (CV) systems are
quantum systems with infinite-dimensional Hilbert spaces like, for instance,
the bosonic modes of a radiation field. In particular, bosonic modes with
Gaussian statistics, i.e., in \emph{Gaussian states} \cite{GaussianStates},
are today extremely important, thanks to their experimental accessibility
and the relative simplicity of their mathematical description.

In the CV framework, the quantum discrimination of Gaussian states can be
seen as a central task. Such a problem was first considered in Ref.~\cite%
{QCbound2}, where a formula for the quantum Chernoff bound has already been
derived. In our paper, we recast this formula by making explicit its
dependence on the symplectic spectra of the involved Gaussian states. The
computational difficulty of this formula relies on the fact that, besides
the symplectic spectra (easy to compute), one must also calculate the
symplectic transformations that diagonalize the corresponding correlation
matrices. The derivation of these symplectic transformations can be in fact
very hard, especially when many bosonic modes are involved in the process.
In order to simplify this computational problem, here we resort to standard
algebraic inequalities, i.e., the Minkowski inequality and the Young's
inequality. Thanks to these inequalities, we can manipulate the formula of
the quantum Chernoff bound and derive much simpler upper bounds for the
discrimination of Gaussian states. These bounds, that we call \emph{%
Minkowski bound} and \emph{Young bound}, are much easier to compute since
they depend on the symplectic spectra only. Notice that, because of this
simplification, these bounds are inevitably weaker than the quantum Chernoff
bound. In particular, they are useful when the Gaussian states are \emph{%
unitarily inequivalent}, i.e., not connected by any unitary transformation
(e.g., displacement, rotation or squeezing). On the one hand, this is surely
a restriction for the general application of our results. On the other hand,
inequivalent Gaussian states arise in many physical situations, and
easy-to-compute upper\ bounds can represent the unique feasible solution
when the number of modes is very high.

The structure of the paper is the following. In Sec.~\ref{IntroSEC} we
review some of the basic notions about Gaussian states, with a special
regard for their normal mode decomposition. In Sec.~\ref{qcbSEC} we review
the quantum Chernoff bound and re-formulate the corresponding expression for
Gaussian states. The subsequent Sec.~\ref{compboundsSEC} contains the
central results of this paper. Here, we derive the computable bounds for
discriminating Gaussian states by combining the quantum Chernoff bound with
the Minkowski determinant inequality and the Young's inequality. We also
provide a simple example in order to compare the various bounds. Sec.~\ref%
{concSEC} is for conclusions.

\section{Gaussian states in a nutshell\label{IntroSEC}}

Let us consider a bosonic system of $n$ modes. This quantum system is
described by a tensor product Hilbert space $\mathcal{H}^{\otimes n}$ and a
vector of quadrature operators $\mathbf{\hat{x}}^{T}:=(\hat{q}_{1},\hat{p}%
_{1},\ldots ,\hat{q}_{n},\hat{p}_{n})$ satisfying the commutation relations%
\begin{equation}
\lbrack \hat{x}_{l},\hat{x}_{m}]=2i\Omega _{lm}~~(1\leq l,m\leq 2n)~,
\label{CCR}
\end{equation}%
where%
\begin{equation}
\mathbf{\Omega }:=\bigoplus\limits_{k=1}^{n}\left(
\begin{array}{cc}
0 & 1 \\
-1 & 0%
\end{array}%
\right)  \label{Symplectic_Form}
\end{equation}%
defines a symplectic form. An arbitrary state of the system is characterized
by a density operator $\rho \in \mathcal{D}(\mathcal{H}^{\otimes n})$ or,
equivalently, by a Wigner representation. In fact, by introducing the Weyl
operator \cite{Weyl}%
\begin{equation}
\hat{D}(\boldsymbol{\xi }):=\exp (i\mathbf{\hat{x}}^{T}\boldsymbol{\xi })~~(%
\boldsymbol{\xi }\in \mathbb{R}^{2n})~,  \label{Weyl operator}
\end{equation}%
an arbitrary $\rho $ is equivalent to a Wigner characteristic function
\begin{equation}
\chi (\boldsymbol{\xi }):=\mathrm{Tr}\left[ \rho \hat{D}(\boldsymbol{\xi })%
\right] ~,  \label{CH_function}
\end{equation}%
or to a Wigner function%
\begin{equation}
W(\mathbf{x}):=\int\limits_{\mathbb{R}^{2n}}\frac{d^{2n}\boldsymbol{\xi }}{%
(2\pi )^{2n}}~\exp \left( -i\mathbf{x}^{T}\boldsymbol{\xi }\right) \chi (%
\boldsymbol{\xi })~.  \label{Wig_function}
\end{equation}%
In the previous Eq.~(\ref{Wig_function}) the continuous variables $\mathbf{x}%
^{T}:=(q_{1},p_{1},\ldots ,q_{n},p_{n})$ are the eigenvalues of $\mathbf{%
\hat{x}}^{T}$. They span the real symplectic space $\mathcal{K}:=(\mathbb{R}%
^{2n},\mathbf{\Omega })$ which is called the \emph{phase space}.

By definition, a bosonic state $\rho $ is called Gaussian if the
corresponding Wigner representation ($\chi $ or $W$) is Gaussian, i.e.,%
\begin{eqnarray}
\chi (\boldsymbol{\xi }) &=&\exp \left[ -\frac{1}{2}\boldsymbol{\xi }^{T}%
\mathbf{V}\boldsymbol{\xi }+i\mathbf{\bar{x}}^{T}\boldsymbol{\xi }\right] ~,
\\
W(\mathbf{x}) &=&\frac{\exp \left[ -\frac{1}{2}(\mathbf{x}-\mathbf{\bar{x}}%
)^{T}\mathbf{V}^{-1}(\mathbf{x}-\mathbf{\bar{x}})\right] }{(2\pi )^{n}\sqrt{%
\det \mathbf{V}}}~.
\end{eqnarray}%
In such a case, the state $\rho $ is fully characterized by its displacement
$\mathbf{\bar{x}:}=\mathrm{Tr}(\mathbf{\hat{x}}\rho )$ and its correlation
matrix (CM) $\mathbf{V}$, with entries%
\begin{equation}
V_{lm}:=\tfrac{1}{2}\mathrm{Tr}\left[ \left\{ \Delta \hat{x}_{l},\Delta \hat{%
x}_{m}\right\} \rho \right] ~,
\end{equation}%
where $\Delta \hat{x}_{l}:=\hat{x}_{l}-\mathrm{Tr}(\hat{x}_{l}\rho )$ and $%
\{,\}$ is the anticommutator. The CM is a $2n\times 2n$, real and symmetric
matrix which must satisfy the uncertainty principle%
\begin{equation}
\mathbf{V}+i\mathbf{\Omega }\geq 0~,  \label{BONA_FIDE}
\end{equation}%
directly coming from Eq.~(\ref{CCR}) and implying $\mathbf{V}>0$.

Fundamental properties of the Gaussian states can be easily expressed via
the symplectic manipulation of their CM's. By definition, a matrix $\mathbf{S%
}$ is called \emph{symplectic} when it preserves the symplectic form of Eq.~(%
\ref{Symplectic_Form}), i.e.,
\begin{equation}
\mathbf{S\Omega S}^{T}=\mathbf{\Omega }~.  \label{Sympl_cond}
\end{equation}%
Then, according to the Williamson's theorem, for every CM $\mathbf{V}$ there
exists a symplectic matrix $\mathbf{S}$ such that
\begin{equation}
\mathbf{V}=\mathbf{\mathbf{S}}\left(
\begin{array}{ccccc}
\nu _{1} &  &  &  &  \\
& \nu _{1} &  &  &  \\
&  & \ddots &  &  \\
&  &  & \nu _{n} &  \\
&  &  &  & \nu _{n}%
\end{array}%
\right) \mathbf{\mathbf{S}}^{T}=\mathbf{\mathbf{S}}\left[ \bigoplus%
\limits_{k=1}^{n}\nu _{k}\mathbf{I}_{k}\right] \mathbf{\mathbf{S}}^{T}~,
\label{William_DEC}
\end{equation}%
where the set $\{\nu _{1},\cdots ,\nu _{n}\}$ is called \emph{symplectic
spectrum} \cite{Spectrum}. In particular, this spectrum satisfies
\begin{equation}
\prod\limits_{k=1}^{n}\nu _{k}=\sqrt{\det \mathbf{V}}~,
\end{equation}%
since $\det \mathbf{\mathbf{S}}=1$. By applying the symplectic
diagonalization of Eq.~(\ref{William_DEC}) to Eq.~(\ref{BONA_FIDE}), one can
write the uncertainty principle in the simpler form \cite{Salerno1}
\begin{equation}
\nu _{k}\geq 1~~\text{and~}~\mathbf{V}>0~.  \label{Heis_spectrum}
\end{equation}

\subsection{Normal mode decomposition of Gaussian states and its application
to power states}

An affine symplectic transformation%
\begin{equation}
(\mathbf{\bar{x}},\mathbf{S}):\mathbf{x\rightarrow Sx+\bar{x}}~,
\label{affine_sympl}
\end{equation}%
acting on the phase space $\mathcal{K}:=(\mathbb{R}^{2n},\mathbf{\Omega })$
results in a simple congruence transformation $\mathbf{V\rightarrow SVS}^{T}$
at the level of the CM. In the space of density operators $\mathcal{D}(%
\mathcal{H}^{\otimes n})$, the transformation of Eq.~(\ref{affine_sympl})
corresponds instead to the transformation
\begin{equation}
\rho \rightarrow \hat{U}_{\mathbf{\bar{x},S}}~\rho ~\hat{U}_{\mathbf{\bar{x}%
,S}}^{\dagger }~,
\end{equation}%
where the unitary $\hat{U}_{\mathbf{\bar{x},S}}=\hat{D}(\mathbf{\bar{x}})%
\hat{U}_{\mathbf{S}}$ is determined by the affine pair $(\mathbf{\bar{x}},%
\mathbf{S})$ and preserves the Gaussian character of the state (Gaussian
unitary). As a consequence, the symplectic diagonalization of Eq.~(\ref%
{William_DEC}) corresponds to a \emph{normal mode decomposition} of the
Gaussian state%
\begin{equation}
\rho =\hat{U}_{\mathbf{\bar{x},S}}\left[ \bigotimes\limits_{k=1}^{n}\sigma
(\nu _{k})\right] \hat{U}_{\mathbf{\bar{x},S}}^{\dagger }~,  \label{Dual}
\end{equation}%
where%
\begin{equation}
\sigma (\nu _{k}):=\frac{2}{\nu _{k}+1}\sum\limits_{j=0}^{\infty }\left(
\frac{\nu _{k}-1}{\nu _{k}+1}\right) ^{j}\left\vert j\right\rangle
_{k}\left\langle j\right\vert  \label{Thermal_state}
\end{equation}%
is a thermal state with mean photon number $\bar{n}_{k}=(\nu _{k}-1)/2$ ($%
\{\left\vert j\right\rangle _{k}\}_{j=0}^{\infty }$ are the number states
for the $k$th mode). Thanks to the normal mode decomposition $(\mathbf{\bar{x%
}},\mathbf{S},\{\nu _{k}\})$ of Eq.~(\ref{Dual}), one can easily compute
every positive power of an $n$-mode Gaussian state $\rho $. In fact, let us
introduce the two basic functions%
\begin{equation}
\Phi _{p}^{\pm }(x):=\left( x+1\right) ^{p}\pm \left( x-1\right) ^{p}~,
\label{Omegas}
\end{equation}%
which are nonnegative for every $x\geq 1$ and $p\geq 0$. Let us also
construct%
\begin{equation}
G_{p}(x):=\frac{2^{p}}{\Phi _{p}^{-}(x)}=\frac{2^{p}}{\left( x+1\right)
^{p}-\left( x-1\right) ^{p}}~,  \label{G_function}
\end{equation}%
and
\begin{equation}
\Lambda _{p}(x):=\frac{\Phi _{p}^{+}(x)}{\Phi _{p}^{-}(x)}=\frac{\left(
x+1\right) ^{p}+\left( x-1\right) ^{p}}{\left( x+1\right) ^{p}-\left(
x-1\right) ^{p}}~.  \label{Lambda_p_Function}
\end{equation}%
Then, we have the following

\begin{lemma}
\label{lemmaDEC}An arbitrary positive power $\rho ^{p}$ of an $n$-mode
Gaussian state $\rho $ can be written as%
\begin{equation}
\rho ^{p}=\left( \mathrm{Tr}\rho ^{p}\right) \rho (p)~,  \label{Rho_power}
\end{equation}%
where%
\begin{equation}
\mathrm{Tr}\rho ^{p}=\prod\limits_{k=1}^{n}G_{p}(\nu _{k})~,
\label{Purities_Symplectic}
\end{equation}%
and%
\begin{equation}
\rho (p):=\hat{U}_{\mathbf{\bar{x},S}}\left\{
\bigotimes\limits_{k=1}^{n}\sigma \left[ \Lambda _{p}(\nu _{k})\right]
\right\} \hat{U}_{\mathbf{\bar{x},S}}^{\dagger }~.  \label{Rho_p}
\end{equation}
\end{lemma}

\bigskip

\noindent \textbf{Proof.}~By setting
\begin{equation}
\nu _{k}=\frac{1+\eta _{k}}{1-\eta _{k}}~\left( \Longleftrightarrow \eta
_{k}=\frac{\nu _{k}-1}{\nu _{k}+1}\right)  \label{ni_and_beta}
\end{equation}%
into Eq.~(\ref{Thermal_state}), we have the following equivalent expression
for the thermal state%
\begin{equation}
\sigma (\eta _{k})=(1-\eta _{k})\sum\limits_{j=0}^{\infty }\eta
_{k}^{j}\left\vert j\right\rangle _{k}\left\langle j\right\vert ~.
\label{Themal_beta}
\end{equation}%
By iterating Eq.~(\ref{Dual}) we get%
\begin{equation}
\rho ^{p}=\hat{U}_{\mathbf{\bar{x},S}}\left\{ \bigotimes\limits_{k=1}^{n}%
\left[ \sigma (\eta _{k})\right] ^{p}\right\} \hat{U}_{\mathbf{\bar{x},S}%
}^{\dagger }  \label{rho_pi_beta}
\end{equation}%
for every $p\geq 0$. Then, from Eq.~(\ref{Themal_beta}), we get%
\begin{equation}
\left[ \sigma (\eta _{k})\right] ^{p}=(1-\eta
_{k})^{p}\sum\limits_{j=0}^{\infty }(\eta _{k}^{p})^{j}\left\vert
j\right\rangle _{k}\left\langle j\right\vert =\frac{(1-\eta _{k})^{p}}{%
1-\eta _{k}^{p}}\sigma (\eta _{k}^{p})~,
\end{equation}%
which, inserted into Eq.~(\ref{rho_pi_beta}), leads to the expression
\begin{equation}
\rho ^{p}=\left[ \prod\limits_{k=1}^{n}\frac{(1-\eta _{k})^{p}}{1-\eta
_{k}^{p}}\right] \left\{ \hat{U}_{\mathbf{\bar{x},S}}\left[
\bigotimes\limits_{k=1}^{n}\sigma (\eta _{k}^{p})\right] \hat{U}_{\mathbf{%
\bar{x},S}}^{\dagger }\right\} ~.  \label{rho_pi_proof}
\end{equation}%
Now, from Eq.~(\ref{rho_pi_proof}) we have%
\begin{equation}
\mathrm{Tr}\rho ^{p}=\prod\limits_{k=1}^{n}\frac{(1-\eta _{k})^{p}}{1-\eta
_{k}^{p}}~,
\end{equation}%
and, applying Eq.~(\ref{ni_and_beta}), we get
\begin{equation}
\mathrm{Tr}\rho ^{p}=\prod\limits_{k=1}^{n}\frac{2^{p}}{\left( \nu
_{k}+1\right) ^{p}-\left( \nu _{k}-1\right) ^{p}}~,
\end{equation}%
which is equivalent to Eqs.~(\ref{Purities_Symplectic}) and~(\ref{G_function}%
). Then, we can easily derive the symplectic eigenvalue $\nu _{k,p}$ of the
thermal state $\sigma (\eta _{k}^{p})$ which is present in Eq.~(\ref%
{rho_pi_proof}). In fact, by using Eq.~(\ref{ni_and_beta}), we get%
\begin{equation}
\nu _{k,p}=\frac{1+\eta _{k}^{p}}{1-\eta _{k}^{p}}=\frac{(\nu
_{k}+1)^{p}+(\nu _{k}-1)^{p}}{(\nu _{k}+1)^{p}-(\nu _{k}-1)^{p}}:=\Lambda
_{p}(\nu _{k})~,  \label{lambda_proof}
\end{equation}%
i.e., $\nu _{k,p}$ is connected to the original eigenvalue $\nu _{k}$ by the
$\Lambda $-function of Eq.~(\ref{Lambda_p_Function}). Finally, by inserting
all the previous results into Eq.~(\ref{rho_pi_proof}) we get the formula of
Eq.~(\ref{Rho_power}). $\blacksquare $

Notice that, thanks to the formula of Eq.~(\ref{Rho_power}), the \emph{%
unnormalized} power state $\rho ^{p}$ is simply expressed in terms of the
symplectic spectrum $\{\nu _{k}\}$ and the affine pair $(\mathbf{\bar{x}},%
\mathbf{S})$ decomposing the original Gaussian state $\rho $ according to
Eq.~(\ref{Dual}). In particular, the CM $\mathbf{V}(p)$\ of the \emph{%
normalized} power state $\rho (p)$ is simply related to the CM $\mathbf{V=V}%
(1)$ of the original state $\rho =\rho (1)$ by the relation%
\begin{equation}
\mathbf{V}(p)=\mathbf{S}\left[ \bigoplus\limits_{k=1}^{n}\Lambda _{p}(\nu
_{k})\mathbf{I}_{k}\right] \mathbf{S}^{T}~.  \label{V_p}
\end{equation}

\section{Quantum Chernoff Bound\label{qcbSEC}}

Let us review the general problem of quantum state discrimination (which we
specialize to Gaussian states of bosonic modes from the next subsection~\ref%
{SubSecFormGauss}). This problem consists in distinguishing between two
possible states $\rho _{A}$ and $\rho _{B}$, which are equiprobable for a
quantum system \cite{Equiprob}. In this discrimination, we suppose that $N$
identical copies of the quantum system are available for a generalized
quantum measurement, i.e., a positive operator valued measure (POVM) \cite%
{NielsenBook}. In other words, we apply a POVM to $N$ copies of the quantum
system in order to choose between two equiprobable hypotheses about its
global state $\rho ^{N}$, i.e.,%
\begin{equation}
H_{A}:\rho ^{N}=\underset{N}{\underbrace{\rho _{A}\otimes \cdots \otimes
\rho _{A}}}:=\rho _{A}^{N}~,  \label{H_0}
\end{equation}%
and%
\begin{equation}
H_{B}:\rho ^{N}=\underset{N}{\underbrace{\rho _{B}\otimes \cdots \otimes
\rho _{B}}}:=\rho _{B}^{N}~.  \label{H_1}
\end{equation}%
In order to achieve an optimal discrimination, it is sufficient to consider
a dichotomic POVM $\{\hat{E}_{A},\hat{E}_{B}\}$, whose Kraus operators $\hat{%
E}_{A}$ and $\hat{E}_{B}$ are associated to the hypotheses $H_{A}$ and $%
H_{B} $, respectively. By performing such a dichotomic POVM $\{\hat{E}_{A},%
\hat{E}_{B}\}$, we get a correct answer up to an error probability%
\begin{gather}
P_{err}^{(N)}=\tfrac{1}{2}P(H_{A}|H_{B})+\tfrac{1}{2}P(H_{B}|H_{A})  \notag
\\
=\tfrac{1}{2}\mathrm{Tr}\left( \hat{E}_{A}\rho _{B}^{N}\right) +\tfrac{1}{2}%
\mathrm{Tr}\left( \hat{E}_{B}\rho _{A}^{N}\right) ~.
\end{gather}%
Clearly, the optimal POVM\ will be the one minimizing $P_{err}^{(N)}$. Now,
since $\hat{E}_{A}=\hat{I}-\hat{E}_{B}$, we can introduce the \emph{Helstrom
matrix} \cite{Helstrom}
\begin{equation}
\gamma :=\rho _{B}^{N}-\rho _{A}^{N}~,
\end{equation}%
and write%
\begin{equation}
P_{err}^{(N)}=\tfrac{1}{2}-\tfrac{1}{2}\mathrm{Tr}\left( \gamma \hat{E}%
_{B}\right) ~.  \label{P_err_Eq1}
\end{equation}%
The error probability of Eq.~(\ref{P_err_Eq1}) can be now minimized over the
Kraus operator $\hat{E}_{B}$ only. Since $\mathrm{Tr}\left( \gamma \right)
=0 $, the Helstrom matrix $\gamma $ has both positive and negative
eigenvalues. As a consequence, the optimal $\hat{E}_{B}$ is the projector
onto the positive part $\gamma _{+}$ of $\gamma $ (i.e., the projector onto
the subspace spanned by the eigenstates of $\gamma $ with positive
eigenvalues). By assuming this optimal operator, we have
\begin{equation}
\mathrm{Tr}\left( \gamma \hat{E}_{B}\right) =\mathrm{Tr}\left( \gamma
_{+}\right) =\frac{1}{2}\left\Vert \gamma \right\Vert _{1}~,
\end{equation}%
where%
\begin{equation}
\left\Vert \gamma \right\Vert _{1}:=\mathrm{Tr}\left\vert \gamma \right\vert
=\mathrm{Tr}\sqrt{\gamma ^{\dagger }\gamma }
\end{equation}%
is the trace norm of the Helstrom matrix $\gamma $. Thus, the \emph{minimal}
error probability $P^{(N)}:=\min P_{err}^{(N)}$ is equal to%
\begin{equation}
P^{(N)}=\tfrac{1}{2}\left( 1-\tfrac{1}{2}\left\Vert \rho _{B}^{N}-\rho
_{A}^{N}\right\Vert _{1}\right) ~.  \label{P_err_Eq2}
\end{equation}

The computation of the trace norm in Eq.~(\ref{P_err_Eq2}) is rather
difficult. Luckily, one can always resort to the quantum Chernoff bound \cite%
{QCbound1}%
\begin{equation}
P^{(N)}\leq P_{QC}^{(N)}~,
\end{equation}%
where%
\begin{equation}
P_{QC}^{(N)}=\frac{1}{2}\exp (-\kappa N)~,
\end{equation}%
and \cite{Infimum}%
\begin{equation}
\kappa :=-\log \left[ \inf_{0\leq s\leq 1}\mathrm{Tr}\left( \rho
_{A}^{s}\rho _{B}^{1-s}\right) \right] ~.  \label{QC_coeff}
\end{equation}%
More simply, this bound can be written as%
\begin{equation}
P_{QC}^{(N)}=\frac{1}{2}\left[ \inf_{0\leq s\leq 1}Q_{s}\right] ^{N}~,
\label{P_QCN}
\end{equation}%
where%
\begin{equation}
Q_{s}:=\mathrm{Tr}\left( \rho _{A}^{s}\rho _{B}^{1-s}\right) ~.
\label{Qs_Eq1}
\end{equation}%
Notice that the quantum Chernoff bound involves a minimization in the
variable $s$. By setting $s=1/2$ in Eq.~(\ref{P_QCN}), one can also define
the quantum version of the Bhattacharyya bound \cite{BHA}
\begin{equation}
P_{B}^{(N)}:=\frac{1}{2}\left[ \mathrm{Tr}\left( \sqrt{\rho _{A}}\sqrt{\rho
_{B}}\right) \right] ^{N}~,
\end{equation}%
which clearly satisfies%
\begin{equation}
P_{QC}^{(N)}\leq P_{B}^{(N)}~.
\end{equation}%
In particular, for $\rho _{A}-\rho _{B}=\delta \rho \simeq 0$, one can show
that $P_{QC}^{(N)}\simeq P_{B}^{(N)}$. Notice that we also have the
following inequalities \cite{QCbound2,NielsenBook}%
\begin{equation}
F_{-}\leq P^{(1)}\leq P_{QC}^{(1)}\leq F_{+}~,  \label{Bounds_Fidelities}
\end{equation}%
where%
\begin{equation}
F_{-}:=\frac{1-\sqrt{1-F(\rho _{A},\rho _{B})}}{2}~,~F_{+}:=\frac{\sqrt{%
F(\rho _{A},\rho _{B})}}{2}~,  \label{FidMP}
\end{equation}%
and%
\begin{equation}
F(\rho _{A},\rho _{B}):=\left[ \mathrm{Tr}\left( \sqrt{\sqrt{\rho _{A}}\rho
_{B}\sqrt{\rho _{A}}}\right) \right] ^{2}  \label{Fidelity_Gen_Formula}
\end{equation}%
is the fidelity between $\rho _{A}$ and $\rho _{B}$ \cite{Fuchs}. In
particular, if one of the two states is pure, e.g., $\rho _{A}=\left\vert
\varphi \right\rangle _{A}\left\langle \varphi \right\vert $, then we simply
have%
\begin{equation}
P_{QC}^{(1)}=\frac{F(\left\vert \varphi \right\rangle _{A}\left\langle
\varphi \right\vert ,\rho _{B})}{2}~.  \label{QC_purecase}
\end{equation}

\subsection{Formula for Gaussian states\label{SubSecFormGauss}}

Let us now specialize the problem of quantum state discrimination to
Gaussian states of $n$ bosonic modes. In this case, the quantum Chernoff
bound can be expressed by a relatively simple formula thanks to the normal
mode decomposition $(\mathbf{\bar{x}},\mathbf{S},\{\nu _{k}\})$ of Eq.~(\ref%
{Dual}).

\begin{theorem}
\label{QC_Gaussian}Let us consider two arbitrary $n$-mode Gaussian states $%
\rho _{A}$ and $\rho _{B}$ with normal mode decompositions $(\mathbf{\bar{x}}%
_{A},\mathbf{S}_{A},\{\alpha _{k}\})$ and $(\mathbf{\bar{x}}_{B},\mathbf{S}%
_{B},\{\beta _{k}\})$. Then, we have%
\begin{equation}
Q_{s}=\bar{Q}_{s}\exp \left\{ -\tfrac{1}{2}\mathbf{d}^{T}\left[ \mathbf{V}%
_{A}(s)+\mathbf{V}_{B}(1-s)\right] ^{-1}\mathbf{d}\right\} ~,  \label{Qs_Eq2}
\end{equation}%
where%
\begin{equation}
\bar{Q}_{s}:=\frac{2^{n}\prod\limits_{k=1}^{n}G_{s}(\alpha
_{k})G_{1-s}(\beta _{k})}{\sqrt{\det \left[ \mathbf{V}_{A}(s)+\mathbf{V}%
_{B}(1-s)\right] }}~.  \label{Qs_Eq3}
\end{equation}%
In these formulas $\mathbf{d}:=\mathbf{\bar{x}}_{A}-\mathbf{\bar{x}}_{B}$ and%
\begin{gather}
\mathbf{V}_{A}(s)=\mathbf{S}_{A}\left[ \bigoplus\limits_{k=1}^{n}\Lambda
_{s}(\alpha _{k})\mathbf{I}_{k}\right] \mathbf{S}_{A}^{T}~,  \label{V_as} \\
\mathbf{V}_{B}(1-s)=\mathbf{S}_{B}\left[ \bigoplus\limits_{k=1}^{n}\Lambda
_{1-s}(\beta _{k})\mathbf{I}_{k}\right] \mathbf{S}_{B}^{T}~.  \label{V_bs}
\end{gather}
\end{theorem}

\noindent \textbf{Proof.}~By applying Lemma~\ref{lemmaDEC} to Eq.~(\ref%
{Qs_Eq1}), we get
\begin{equation}
Q_{s}=\mathcal{N}\mathrm{Tr}\left[ \rho _{A}(s)\rho _{B}(1-s)\right] ~,
\label{Qs_Eq4}
\end{equation}%
where%
\begin{equation}
\mathcal{N}:=\left( \mathrm{Tr}\rho _{A}^{s}\right) \left( \mathrm{Tr}\rho
_{B}^{1-s}\right) =\prod\limits_{k=1}^{n}G_{s}(\alpha _{k})G_{1-s}(\beta
_{k})~,  \label{Norm_Factor}
\end{equation}%
and $\rho _{A}(s),\rho _{B}(1-s)$ are two Gaussian states defined according
to Eq.~(\ref{Rho_p}). In particular, the CM's of these states are given by
Eqs.~(\ref{V_as}) and~(\ref{V_bs})\ [according to Eq.~(\ref{V_p})]. For an
arbitrary pair of $n$-mode Gaussian states $\rho ,\rho ^{\prime }$ with
characteristic functions $\chi ,\chi ^{\prime }$ and moments $(\mathbf{V},%
\mathbf{\bar{x}})$ and $(\mathbf{V}^{\prime },\mathbf{\bar{x}}^{\prime })$,
we have the trace rule%
\begin{gather}
\mathrm{Tr}\left( \rho \rho ^{\prime }\right) =\int\limits_{\mathbb{R}^{2n}}%
\frac{d^{2n}\boldsymbol{\xi }}{\pi ^{n}}~\chi (\boldsymbol{\xi })\chi
^{\prime }(-\boldsymbol{\xi })  \notag \\
=2^{n}\frac{\exp \left[ -\tfrac{1}{2}(\mathbf{\bar{x}-\bar{x}}^{\prime
})^{T}\left( \mathbf{V}+\mathbf{V}^{\prime }\right) ^{-1}(\mathbf{\bar{x}-%
\bar{x}}^{\prime })\right] }{\sqrt{\det \left( \mathbf{V}+\mathbf{V}^{\prime
}\right) }}~.  \label{Trace_into_Det}
\end{gather}%
Then, by using Eq.~(\ref{Trace_into_Det}) into Eq.~(\ref{Qs_Eq4}), we easily
get Eqs.~(\ref{Qs_Eq2}) and~(\ref{Qs_Eq3}).$~\blacksquare $

Thanks to the previous theorem, the Chernoff quantity $Q_{s}$ can be
directly computed from the normal mode decompositions $(\mathbf{\bar{x}}_{A},%
\mathbf{S}_{A},\{\alpha _{k}\})$ and $(\mathbf{\bar{x}}_{B},\mathbf{S}%
_{B},\{\beta _{k}\})$ of the Gaussian states. Notice that this theorem is
already contained in Ref.~\cite{QCbound2} but here the formula of Eqs.~(\ref%
{Qs_Eq2}) and~(\ref{Qs_Eq3}) is conveniently expressed in terms of the
symplectic spectra $\{\alpha _{k}\}$ and $\{\beta _{k}\}$.

In applying this theorem, the more difficult task is the algebraic
computation of the symplectic matrices $\mathbf{S}_{A}$ and $\mathbf{S}_{B}$
to be used in Eqs.~(\ref{V_as}) and~(\ref{V_bs}). In fact, while finding the
symplectic eigenvalues $\{\nu _{k}\}$ is relatively easy (since they are the
degenerate solutions of a $2n$-degree polynomial), finding the diagonalizing
symplectic matrix $\mathbf{S}$ is computationally harder (since it
corresponds to the construction of a symplectic basis \cite{Arnold}). For
this reason, it is very helpful to derive bounds for the minimal error
probability $P^{(N)}$ which do not depend on $\mathbf{S}$\ and, therefore,
are much easier to compute.

\section{Computable bounds for discriminating Gaussian states\label%
{compboundsSEC}}

Let us derive bounds that do not depend on the affine symplectic
transformations $(\mathbf{\bar{x}}_{A},\mathbf{S}_{A})$ and $(\mathbf{\bar{x}%
}_{B},\mathbf{S}_{B})$, but only on the symplectic spectra $\{\alpha _{k}\}$
and $\{\beta _{k}\}$. This is possible by simplifying the determinant in
Eq.~(\ref{Qs_Eq3}) involving the two positive matrices $\mathbf{V}_{A}(s)$
and $\mathbf{V}_{B}(1-s)$. Such a determinant can be decomposed into a sum
of determinants by resorting to the Minkowski determinant inequality \cite%
{Bhatia}. In general, such an algebraic theorem is valid for non-negative
complex matrices in any dimension (see, e.g., Appendix~\ref{MinkDetIneq_APP}%
). In particular, it can be specialized to positive real matrices in even
dimension and, therefore, to correlation matrices.

\begin{lemma}
\label{ALGlemma}Let us consider a pair of $2n\times 2n$ real, symmetric and
positive matrices $\mathbf{K}$ and $\mathbf{L}$. Then, we have the Minkowski
determinant inequality
\begin{equation}
\left[ \det \left( \mathbf{K+L}\right) \right] ^{1/2n}\geq \left( \det
\mathbf{K}\right) ^{1/2n}+\left( \det \mathbf{L}\right) ^{1/2n}~.
\label{Mink_ineq}
\end{equation}
\end{lemma}

\noindent By combining Theorem~\ref{QC_Gaussian} and Lemma~\ref{ALGlemma},
we can prove the following

\begin{theorem}
\label{BoundsTHEO}Let us consider two arbitrary $n$-mode Gaussian states $%
\rho _{A}$ and $\rho _{B}$ with symplectic spectra $\{\alpha _{k}\}$ and $%
\{\beta _{k}\}$. Then, we have the \textquotedblleft Minkowski
bound\textquotedblright
\begin{equation}
P_{QC}^{(N)}\leq M^{(N)}:=\frac{1}{2}\left[ \inf_{0\leq s\leq 1}M_{s}\right]
^{N},  \label{Minkowski_Bound}
\end{equation}%
where%
\begin{equation}
M_{s}:=4^{n}\left[ \prod\limits_{k=1}^{n}\Psi _{s}(\alpha _{k},\beta
_{k})+\prod\limits_{k=1}^{n}\Psi _{1-s}(\beta _{k},\alpha _{k})\right] ^{-n},
\label{Sum_bound}
\end{equation}%
and%
\begin{equation}
\Psi _{p}(x,y):=\left[ \Phi _{p}^{+}(x)\Phi _{1-p}^{-}(y)\right] ^{1/n}.
\label{PSI}
\end{equation}
\end{theorem}

\noindent \textbf{Proof.}~By taking the $n$th power of Eq.~(\ref{Mink_ineq}%
), we get%
\begin{equation}
\left[ \det \left( \mathbf{K+L}\right) \right] ^{1/2}\geq \left[ \left( \det
\mathbf{K}\right) ^{1/2n}+\left( \det \mathbf{L}\right) ^{1/2n}\right] ^{n}~.
\label{SumDets}
\end{equation}%
Such inequality can be directly applied to the CM's $\mathbf{V}_{A}(s)$ and $%
\mathbf{V}_{B}(1-s)$ of Eqs.~(\ref{V_as}) and~(\ref{V_bs}). Then, by
inserting the result into Eq.~(\ref{Qs_Eq3}), we get%
\begin{equation}
\bar{Q}_{s}\leq \frac{2^{n}\prod\limits_{k=1}^{n}G_{s}(\alpha
_{k})G_{1-s}(\beta _{k})}{\left\{ \left[ \det \mathbf{V}_{A}(s)\right] ^{%
\frac{1}{2n}}+\left[ \det \mathbf{V}_{B}(1-s)\right] ^{\frac{1}{2n}}\right\}
^{n}}:=M_{s}~.  \label{M_s_proof}
\end{equation}%
By using the binomial expansion and the relations%
\begin{gather}
\det \mathbf{V}_{A}(s)=\prod\limits_{k=1}^{n}\left[ \Lambda _{s}(\alpha _{k})%
\right] ^{2}~,  \label{det_V0} \\
\det \mathbf{V}_{B}(1-s)=\prod\limits_{k=1}^{n}\left[ \Lambda _{1-s}(\beta
_{k})\right] ^{2}~,  \label{det_V1}
\end{gather}%
we get%
\begin{equation}
M_{s}^{-1}=\frac{1}{2^{n}}\sum_{i=0}^{n}\dbinom{n}{i}\prod\limits_{k=1}^{n}%
\frac{\left[ \Lambda _{s}(\alpha _{k})\right] ^{\frac{i}{n}}\left[ \Lambda
_{1-s}(\beta _{k})\right] ^{\frac{n-i}{n}}}{G_{s}(\alpha _{k})G_{1-s}(\beta
_{k})}~.
\end{equation}%
Now, by using the Eqs.~(\ref{G_function}) and~(\ref{Lambda_p_Function}), we
get%
\begin{gather}
M_{s}^{-1}=\frac{1}{4^{n}}\sum_{i=0}^{n}\dbinom{n}{i}\left\{ \left[
\prod\limits_{k=1}^{n}\Phi _{s}^{+}(\alpha _{k})\Phi _{1-s}^{-}(\beta _{k})%
\right] ^{\frac{i}{n}}\right.  \notag \\
\left. \times \left[ \prod\limits_{k=1}^{n}\Phi _{s}^{-}(\alpha _{k})\Phi
_{1-s}^{+}(\beta _{k})\right] ^{\frac{n-i}{n}}\right\} ~,
\end{gather}%
and, by using Eq.~(\ref{PSI}), we derive%
\begin{gather}
M_{s}^{-1}=\frac{1}{4^{n}}\sum_{i=0}^{n}\binom{n}{i}\left\{ \left[
\prod\limits_{k=1}^{n}\Psi _{s}(\alpha _{k},\beta _{k})\right] ^{i}\right.
\notag \\
\left. \times \left[ \prod\limits_{k=1}^{n}\Psi _{1-s}(\beta _{k},\alpha
_{k})\right] ^{n-i}\right\}  \notag \\
=\frac{1}{4^{n}}\left[ \prod\limits_{k=1}^{n}\Psi _{s}(\alpha _{k},\beta
_{k})+\prod\limits_{k=1}^{n}\Psi _{1-s}(\beta _{k},\alpha _{k})\right] ^{n}~.
\end{gather}%
Latter quantity corresponds to the inverse of the one in Eq.~(\ref{Sum_bound}%
). Now, since every convex combination of positive matrices is still
positive, we have that $\exp \{\cdots \}\leq 1$ in Eq.~(\ref{Qs_Eq2}). Then,
we get%
\begin{equation}
Q_{s}\leq \bar{Q}_{s}\leq M_{s}~,
\end{equation}%
leading to the final result of Eq.~(\ref{Minkowski_Bound}).$~\blacksquare $

The basic algebraic property which has been exploited in Theorem~\ref%
{BoundsTHEO} is the concavity of the function \textquotedblleft $\sqrt[2n]{%
\det }$\textquotedblright\ on every convex combination of $2n\times 2n$
positive matrices (like the correlation matrices). Such a property is simply
expressed by the Eq.~(\ref{Mink_ineq}) of Lemma~\ref{ALGlemma}. It enables
us to decompose the determinant of a sum into a sum of determinants and,
therefore, to derive the bound in the \textquotedblleft sum
form\textquotedblright\ of Eq.~(\ref{Sum_bound}). Now, thanks to the Young's
inequality \cite{Young}, every convex combination of positive numbers is
lower bounded by a product of their powers, i.e., for every $k,l>0$ and $%
0\leq \theta \leq 1$, one has%
\begin{equation}
\theta k+(1-\theta )l\geq k^{\theta }l^{1-\theta }~.  \label{Y_ineq}
\end{equation}%
As a consequence, every sum of positive determinants can be bounded by a
product of determinants. Then, by applying the Young's inequality to Theorem~%
\ref{BoundsTHEO}, we can easily derive a weaker bound which is in a
\textquotedblleft product form\textquotedblright . This is shown in the
following corollary. Notice that this bound can be equivalently found by
exploiting the concavity of the function \textquotedblleft $\log \det $%
\textquotedblright\ on every convex combination of positive matrices. (see
Appendix~\ref{MinkDetIneq_APP} for details.)

\begin{corollary}
\label{Bounds_Corollary}Let us consider two arbitrary $n$-mode Gaussian
states $\rho _{A}$ and $\rho _{B}$ with symplectic spectra $\{\alpha _{k}\}$
and $\{\beta _{k}\}$. Then, we have the \textquotedblleft Young
bound\textquotedblright
\begin{equation}
M^{(N)}\leq Y^{(N)}:=\frac{1}{2}\left[ \inf_{0\leq s\leq 1}Y_{s}\right]
^{N}~,  \label{Young_Bound}
\end{equation}%
where%
\begin{equation}
Y_{s}:=2^{n}\prod\limits_{k=1}^{n}\Gamma _{s}(\alpha _{k})\Gamma
_{1-s}(\beta _{k})~,  \label{Product_bound}
\end{equation}%
and%
\begin{equation}
\Gamma _{p}(x):=\left[ (x+1)^{2p}-(x-1)^{2p}\right] ^{-\frac{1}{2}}~.
\label{GAMMA}
\end{equation}
\end{corollary}

\noindent \textbf{Proof.~}From Eq.~(\ref{Y_ineq}) with $\theta =1/2$, we
have that every pair of real numbers $k,l>0$ satisfies
\begin{equation}
k+l\geq 2\sqrt{kl}~.  \label{Young_cons}
\end{equation}%
Then, for positive $\mathbf{K}$ and $\mathbf{L}$, we can apply Eq.~(\ref%
{Young_cons}) with
\begin{equation}
k:=\left( \det \mathbf{K}\right) ^{1/2n}>0~,~l:=\left( \det \mathbf{L}%
\right) ^{1/2n}>0~.
\end{equation}%
This leads to the further lower bound%
\begin{equation}
\left[ \left( \det \mathbf{K}\right) ^{1/2n}+\left( \det \mathbf{L}\right)
^{1/2n}\right] ^{n}\geq 2^{n}\left[ \det \mathbf{K}\det \mathbf{L}\right]
^{1/4}~.  \label{Det_Corollary}
\end{equation}%
By applying Eq.~(\ref{Det_Corollary}) to the CM's $\mathbf{V}_{A}(s)$ and $%
\mathbf{V}_{B}(1-s)$, and inserting the result into Eq.~(\ref{M_s_proof}),
we get%
\begin{equation}
M_{s}\leq \prod\limits_{k=1}^{n}\frac{G_{s}(\alpha _{k})G_{1-s}(\beta _{k})}{%
\sqrt[4]{\det \mathbf{V}_{A}(s)\det \mathbf{V}_{B}(1-s)}}:=Y_{s}~.
\end{equation}%
Then, by using Eqs.~(\ref{det_V0}) and~(\ref{det_V1}), we can write%
\begin{equation}
Y_{s}=\prod\limits_{k=1}^{n}\frac{G_{s}(\alpha _{k})G_{1-s}(\beta _{k})}{%
\sqrt{\Lambda _{s}(\alpha _{k})}\sqrt{\Lambda _{1-s}(\beta _{k})}}~.
\end{equation}%
Exploiting Eqs.~(\ref{G_function}) and~(\ref{Lambda_p_Function}), we can
easily derive Eq.~(\ref{Product_bound}), where%
\begin{equation}
\Gamma _{p}(x):=\left[ \Phi _{p}^{+}(x)\Phi _{p}^{-}(x)\right] ^{-\frac{1}{2}%
}~,
\end{equation}%
also equivalent to Eq.~(\ref{GAMMA}). Finally, since $M_{s}\leq Y_{s}$, the
result of Eq.~(\ref{Young_Bound}) is straightforward.$~\blacksquare $

As stated in Theorem~\ref{BoundsTHEO} and Corollary~\ref{Bounds_Corollary},
the two bounds $M^{(N)}$ and $Y^{(N)}$\ depend only on the symplectic
spectra $\{\alpha _{k}\}$ and $\{\beta _{k}\}$ of the input states $\rho
_{A} $ and $\rho _{B}$. As a consequence, such bounds are useful in
discriminating Gaussian states which are unitarily \emph{inequivalent},
i.e., such that
\begin{equation}
\rho _{A}\neq \hat{U}\rho _{B}\hat{U}^{\dagger }~,
\end{equation}%
for every unitary $\hat{U}$. In fact, since $\rho _{A}$ and $\rho _{B}$ are
Gaussian states, every unitary $\hat{U}$ such that $\rho _{A}=\hat{U}\rho
_{B}\hat{U}^{\dagger }$ must be a Gaussian unitary $\hat{U}=\hat{U}_{\mathbf{%
\bar{x}},\mathbf{S}}$. Its action corresponds therefore to an affine
symplectic trasformation $(\mathbf{\bar{x}},\mathbf{S})$, which cannot
change the symplectic spectrum (so that $\{\alpha _{k}\}=\{\beta _{k}\}$).
Roughly speaking, the previous bounds are useful when the main difference
between $\rho _{A}$ and $\rho _{B}$ is due to the noise, whose variations
break the equivalence and are stored in the symplectic spectra. This
situation is common in several quantum scenarios. For instance, in quantum
illumination \cite{QIll,QIll2}, where two different thermal-noise channels
must be discriminated, or in quantum cloning, when the outputs of an
asymmetric Gaussian cloner \cite{Cerf} must be distinguished.

\subsection{Discrimination of single mode Gaussian states: an example}

\begin{figure}[tbph]
\vspace{+0.2cm}
\par
\begin{center}
\includegraphics[width=0.4\textwidth] {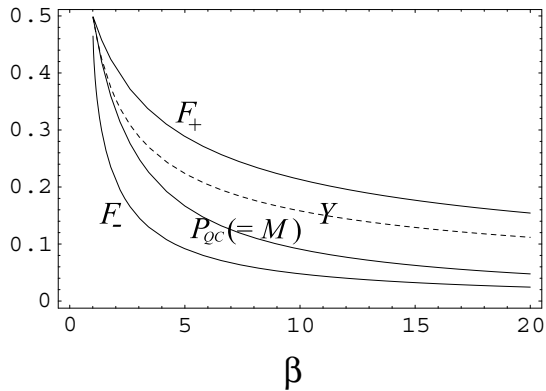}
\end{center}
\par
\vspace{-0.5cm}
\caption{Behavior of the various bounds $Y^{(1)}$, $M^{(1)}$, $P_{QC}^{(1)}$%
, $F_{+}$ and $F_{-}$ versus the eigenvalue $\protect\beta $ in the
discrimination of a thermal state $\protect\sigma (\protect\beta )$ from a
vacuum state. Notice that $M^{(1)}=P_{QC}^{(1)}$ in this example.}
\label{PlotBounds}
\end{figure}

Let us compare the bounds of Theorem~\ref{BoundsTHEO} and Corollary~\ref%
{Bounds_Corollary} with the fidelity bounds of Eq.~(\ref{Bounds_Fidelities})
in a simple case. According to Ref.~\cite{FidFormulas}, the fidelity between
two single-mode Gaussian states $\rho _{A}$ and $\rho _{B}$, with moments $(%
\mathbf{V}_{A},\mathbf{\bar{x}}_{A})$ and $(\mathbf{V}_{B},\mathbf{\bar{x}}%
_{B})$, is given by the formula%
\begin{equation}
F(\rho _{A},\rho _{B})=\frac{2}{\sqrt{\Delta +\delta }-\sqrt{\delta }}\exp %
\left[ -\tfrac{1}{2}\mathbf{d}^{T}(\mathbf{V}_{A}+\mathbf{V}_{B})^{-1}%
\mathbf{d}\right] ,  \label{Fidelity_pairGaussian}
\end{equation}%
where%
\begin{equation}
\Delta :=\det (\mathbf{V}_{A}+\mathbf{V}_{B})~,~\delta :=(\det \mathbf{V}%
_{A}-1)(\det \mathbf{V}_{B}-1)~,  \label{DeltasFORfidelity}
\end{equation}%
and $\mathbf{d}:=\mathbf{\bar{x}}_{A}-\mathbf{\bar{x}}_{B}$. Let us
discriminate between the two single-mode states: $\rho _{A}=\sigma
(1)=\left\vert 0\right\rangle \left\langle 0\right\vert $ (vacuum state) and
$\rho _{B}=\sigma (\beta )$ (arbitrary thermal state). In such a case, it is
very easy to compute the infima of $M_{s}$\ and $Y_{s}$ in Eqs.~(\ref%
{Minkowski_Bound}) and~(\ref{Young_Bound}), respectively. In fact, by
exploiting%
\begin{equation}
\Phi _{p}^{\pm }(1)=\left[ \Gamma _{p}(1)\right] ^{-1}=2^{p}~,~\Phi
_{p}^{+}(x)+\Phi _{p}^{-}(x)=2\left( x+1\right) ^{p}~,
\end{equation}%
we get%
\begin{equation}
M_{s}=\left( \frac{2}{1+\beta }\right) ^{1-s},~Y_{s}=\frac{2^{s}}{\sqrt{%
(\beta +1)^{2s}-(\beta -1)^{2s}}}~,
\end{equation}%
whose infima are taken at $s=0$ and $s=1$, respectively. As a consequence,
for a single copy of the state, we have
\begin{equation}
M^{(1)}=(1+\beta )^{-1}~,~Y^{(1)}=\frac{1}{2\sqrt{\beta }}~.
\end{equation}%
At the same time, we have
\begin{equation}
F(\rho _{A},\rho _{B})=2(1+\beta )^{-1}~,
\end{equation}%
which implies%
\begin{equation}
F_{-}=\frac{1}{2}-\frac{1}{2}\sqrt{\frac{\beta -1}{\beta +1}}~,~F_{+}=\frac{1%
}{\sqrt{2(1+\beta )}}~.
\end{equation}%
By using Eq.~(\ref{QC_purecase}), we also derive%
\begin{equation}
P_{QC}^{(1)}=(1+\beta )^{-1}~.
\end{equation}%
As evident from Fig.~\ref{PlotBounds}, the Young bound $Y^{(1)}$ is tighter
than the fidelity bound $F_{+}$, while the Minkowski bound $M^{(1)}$
coincides exactly with $P_{QC}^{(1)}$ in this case.

\section{Conclusion\label{concSEC}}

We have considered the general problem of discriminating two Gaussian states
of $n$ bosonic modes, supposing that $N$ copies of the state are provided.
To face this problem, we have suitably recasted the formula for the quantum
Chernoff bound given in Ref.~\cite{QCbound2}. By combining this formula with
classical algebraic inequalities (Minkowski determinant inequality and
Young's inequality) we have derived easy-to-compute upper bounds whose
computational hardness is equivalent to finding the symplectic eigenvalues
of the involved Gaussian states. Since these upper bounds depend only on the
symplectic spectra, they are useful in distinguishing Gaussian states which
are unitarily inequivalent. This is indeed a common situation in various
quantum scenarios where the noise is the key element to be discriminated.
For instance, the discrimination between two different thermal-noise
channels is a basic process in quantum sensing and imaging, where nearly
transparent objects must be detected \cite{QIll,QIll2}. Potential
applications of our results concern also quantum cryptography, where the
presence of noise is related to the presence of a malicious eavesdropper.

\section{Acknowledgements}

S.P. was supported by a Marie Curie Outgoing International Fellowship within
the 6th European Community Framework Programme. S.L. was supported by the
W.M. Keck center for extreme quantum information processing (xQIT).

\appendix

\section{Basic algebraic inequalities\label{MinkDetIneq_APP}}

For completeness we report some of the basic algebraic tools used in our
derivation (see also Refs.~\cite{MarcusMink,Bhatia,Horn}). Here we denote by
$\mathbb{M}(m,\mathbb{C})$ the set of $m\times m$ matrices with complex
entries.

\begin{theorem}[Minkowski determinant inequality]
\label{Mink_APP}Let us consider $\mathbf{K,L}\in \mathbb{M}(m,\mathbb{C})$
such that $\mathbf{K}^{\dagger }=\mathbf{K}\geq 0$ and $\mathbf{L}^{\dagger
}=\mathbf{L}\geq 0$. Then%
\begin{equation}
\left[ \det (\mathbf{K+L})\right] ^{\frac{1}{m}}\geq \left( \det \mathbf{K}%
\right) ^{\frac{1}{m}}+\left( \det \mathbf{L}\right) ^{\frac{1}{m}}~.
\label{Mink_main}
\end{equation}%
More generally%
\begin{equation}
\left\{ \det \left[ \theta \mathbf{K+}(1-\theta )\mathbf{L}\right] \right\}
^{\frac{1}{m}}\geq \theta \left( \det \mathbf{K}\right) ^{\frac{1}{m}%
}+(1-\theta )\left( \det \mathbf{L}\right) ^{\frac{1}{m}},
\label{Mink_Theta}
\end{equation}%
for every $0\leq \theta \leq 1$.
\end{theorem}

\noindent By taking the $m$th power of Eq.~(\ref{Mink_main}), it is trivial
to check that%
\begin{equation}
\det (\mathbf{K+L})\geq \det \mathbf{K}+\det \mathbf{L~.}
\end{equation}%
Then, by using the Young's inequality of Eq.~(\ref{Y_ineq}), we can easily
prove the following

\begin{corollary}
\label{Cor_Mink_APP}Let us consider $\mathbf{K,L}\in \mathbb{M}(m,\mathbb{C}%
) $ such that $\mathbf{K}^{\dagger }=\mathbf{K}>0$ and $\mathbf{L}^{\dagger
}=\mathbf{L}>0$. Then%
\begin{equation}
\det \left[ \theta \mathbf{K+(}1-\theta \mathbf{)L}\right] \geq \left( \det
\mathbf{K}\right) ^{\theta }\left( \det \mathbf{L}\right) ^{1-\theta }~,
\label{corollary_Mink}
\end{equation}%
for every $0\leq \theta \leq 1$.
\end{corollary}

\noindent \textbf{Proof.}~By setting $k:=\left( \det \mathbf{K}\right)
^{1/m}>0$ and $l:=\left( \det \mathbf{L}\right) ^{1/m}>0$, we can apply Eq.~(%
\ref{Y_ineq}) to the right hand side of Eq.~(\ref{Mink_Theta}). Then, we get
the result of Eq.~(\ref{corollary_Mink}) by taking the $m$th power.~$%
\blacksquare $

\noindent Notice that, by taking the logarithm of Eq.~(\ref{corollary_Mink}%
), we get%
\begin{equation}
f\left[ \theta \mathbf{K+(}1-\theta \mathbf{)L}\right] \geq \theta f\left(
\mathbf{K}\right) +\mathbf{(}1-\theta \mathbf{)}f\left( \mathbf{L}\right) ~,
\end{equation}%
where
\begin{equation}
f(\mathbf{M}):=\log \det (\mathbf{M})~.
\end{equation}%
In other words, the Corollary~\ref{Cor_Mink_APP} states that the function
\textquotedblleft $\log \det $\textquotedblright\ is concave on convex
combinations of positive matrices. The Theorem~\ref{Mink_APP} instead states
that the function \textquotedblleft $\sqrt[m]{\det }$\textquotedblright\ is
concave on convex combinations of $m$-square non-negative matrices.


\begin{thebibliography}{99}
\bibitem{Weyl} Notice that our choice of the Weyl operator is different from
$\hat{D}(\boldsymbol{\xi }):=\exp (i\mathbf{\hat{x}}^{T}\mathbf{\Omega }%
\boldsymbol{\xi })$, whose definition directly comes from the standard
displacement operator of quantum optics. The choice of Eq.~(\ref{Weyl
operator}) avoids the presence of $\mathbf{\Omega }$ in the characteristic
functions.
\bibitem{Spectrum} The symplectic eigenvalues of $\mathbf{V}$\ can also be
computed as the `standard' eigenvalues of the matrix $|i\mathbf{\Omega V}|$
where the modulus must be understood in the operatorial sense. In fact, the
matrix $i\mathbf{\Omega V}$ is Hermitian and, therefore, diagonalizable by a
unitary transformation. Then, by taking the modulus of its $2n$ real
eigenvalues, one gets the $n$ symplectic eigenvalues of $\mathbf{V}$.
\bibitem{Equiprob} More generally, the problem can be formulated for quantum
states which are not equiprobable.
\bibitem{Infimum} Notice that the infimum instead of the minimum is used in
this formula since a discontinuity may affect the border points $s=0$ and $%
s=1$ \cite{QCbound1,QCbound2}.
\end{thebibliography}
\end{document}